%% file: main.tex
\documentclass[10pt]{article}
\usepackage[letterpaper,margin=0.82in]{geometry}
\usepackage[utf8]{inputenc}
\usepackage[T1]{fontenc}
\usepackage{newtxtext}
\usepackage{amsmath,amssymb}
\usepackage{booktabs}
\usepackage{array}
\usepackage{graphicx}
\PassOptionsToPackage{hyphens}{url}
\usepackage[round,authoryear]{natbib}
\usepackage[hidelinks]{hyperref}
\usepackage{microtype}
\usepackage{placeins}
\newcommand{\Prob}{\mathbb{P}}
\newcommand{\DeepSeek}{DeepSeek-V4-Pro}
\newcommand{\Grok}{Grok-4.3}
\newcommand{\Mistral}{Mistral-medium-3.5}
\newcommand{\GPT}{GPT-5.5/Codex}

\title{When Policies Change Probabilities:\\Modular Decision-Making for LLM Code Review}
\author{\normalsize Rasvik Kudum, Max Corbett, Hitansh Paliwal, Romaisa Fatima, Thomas Jiralerspong, and Sneheel Sarangi}
\date{July 2026}

\hypersetup{
  pdftitle={When Policies Change Probabilities: Modular Decision-Making for LLM Code Review},
  pdfauthor={Rasvik Kudum, Max Corbett, Hitansh Paliwal, Romaisa Fatima, Thomas Jiralerspong, Sneheel Sarangi}
}

\begin{document}
\maketitle

\begin{abstract}
LLM code reviewers often estimate patch risk and make approval decisions in one prompt. A probability should depend on evidence; costs should determine the action taken from it. We test whether four deployed reviewer interfaces preserve this separation using 15,792 responses on 720 candidate patches, with one that passed and one that failed an archived test harness for each of 360 repository issues. In matched calls with the patch and monitor evidence fixed, replacing an equal-cost policy with a 10:1 false-accept policy changes reported failure probabilities by 13.6--16.9 percentage points on average. For every reviewer, the actions returned under the high-cost prompt are worse than rejecting all patches. Applying the same high-cost rule to probabilities elicited under equal costs reduces loss for all four systems, showing that probability elicitation itself contributes to the excess loss. We also evaluate a modular pipeline that elicits risk without policy information, combines an independent monitor score, and applies costs in code. Relative to calibrated reviewer-only scores, the pipeline improves average probability accuracy and, at equal costs, reduces mean loss by .073 per issue while accepting 58--68\% of patches. At 10:1, it accepts none and matches reject-all. Downstream policy can therefore change the probability it is meant to use, motivating separate evaluation of risk, outside evidence, and action.
\end{abstract}

\begin{center}
\small Code and data: \url{https://github.com/rasvik/when-policies-change-probabilities} (commit \texttt{9ece24e}).
\end{center}

\section{Introduction}

Suppose a reviewer assigns a 20\% chance that a patch will fail. An organization then decides that approving a bad patch is ten times more costly than rejecting a good one. The approval threshold should change, but the 20\% estimate should not: the patch and the evidence are unchanged.

LLM-based coding systems increasingly use reviewer models to approve, reject, revise, or rank generated patches \citep{wang2026swereview,zhang2026crab,adams2026radar}. Best-of-$N$ systems use verifier scores to choose among candidate solutions \citep{kwok2026verifier,landesberg2026bestofn}. Such scores may be calibrated, cached, compared across model versions, or passed to teams with different error costs. These uses assume that a score describes the patch rather than the policy that requested it.

Prior work has shown that language-model confidence depends on how it is elicited and has studied whether confidence agrees with actions or whether judge verdicts remain stable under policy reformulations \citep{kadavath2022know,wang2026riskeval,yamin2026belief,weng2026policy}. We ask a different systems question: when the patch and evidence are fixed, can a downstream decision policy change the numeric failure probability itself? Executable patch outcomes let us measure both the probability error and the resulting decision loss.

We study four deployed reviewer interfaces on 720 patches from 360 repository issues. A lower stated failure rate serves as a positive control because it should change the probability. The main intervention replaces an equal-cost prompt with a two-line block that gives a 10:1 false-accept penalty and its derived threshold, while holding the reviewer, patch, code context, and monitor evidence fixed. A second comparison applies one high-cost rule to probabilities from both prompts. Finally, we evaluate a modular pipeline: the reviewer reports risk without policy information, a second model (the \emph{monitor}) supplies an independent failure score, and code combines the scores and applies the decision rule.

The study makes three contributions:
\begin{itemize}
    \item \textbf{Policy dependence.} The joint cost-and-threshold block changes matched probabilities by 13.6--16.9 percentage points on average and worsens average probability error. The change exceeds same-prompt repeat variation for three of the four systems.
    \item \textbf{Decision impact.} For every reviewer, actions returned by the high-cost prompt are worse than rejecting all patches on the balanced benchmark. Under the same high-cost rule, matched equal-cost probabilities reduce loss for all four systems.
    \item \textbf{Modular evaluation.} Relative to calibrated reviewer-only scores, a policy-free reviewer score, a separate monitor score, and a coded controller improve average probability accuracy and equal-cost decision loss. The monitor largely replaces weaker reviewers and only weakly complements the strongest; at 10:1, the pipeline accepts no patches.
\end{itemize}

These claims concern four fixed reviewer interfaces and an archived harness outcome. They do not imply universal behavior or that separate calls are the only workable design.

\section{Related Work}

\subsection{Confidence Reports and Decision Policy}

Language models can report confidence, but the numbers depend on how they are elicited and measured \citep{kadavath2022know,cox2025mapping,yang2026verbalized,kim2026protocol}. Sample consistency and post-hoc calibration can improve their accuracy \citep{lyu2025sample,guo2017calibration}. The closest work studies how confidence and action relate. RiskEval varies error penalties and asks whether abstention follows verbal confidence; Yamin et al. test whether probabilities and actions are jointly coherent; Weng et al. test whether safety-judge verdicts remain stable under policy reformulations \citep{wang2026riskeval,yamin2026belief,weng2026policy}. We ask the complementary question: with the patch and evidence fixed, does the downstream cost-and-threshold rule alter the reported probability itself? Executable outcomes let us measure both probability error and decision loss.

\subsection{Executable Evaluation for Software Agents}

SWE-bench introduced repository issues scored by executable tests, and SWE-rebench automated task collection while adding contamination controls \citep{jimenez2024swebench,badertdinov2025swerebench}. Recent code-review benchmarks and deployed systems ask LLMs to accept, reject, or prioritize patches \citep{zhang2026crab,wang2026swereview,adams2026radar}. We use the executable outcome for a different purpose: as a fixed label in a matched audit of numeric risk reports. This turns a prompt-induced probability change into a measurable change in probability quality and cost-sensitive loss.

\subsection{Monitoring and Modular Decision Systems}

Monitoring work studies whether one model can detect failures or misconduct in another agent and identify cases for escalation \citep{kale2025monitoring}. Work on uncertainty propagation and agent orchestration asks how probability estimates should be carried into later decisions; Calibrate-Then-Act uses calibrated priors to guide cost-aware behavior \citep{papamarkou2026bayes,xia2026uncertainty,ding2026calibrate}. We test two practical questions: whether the monitor adds information beyond the reviewer, and whether a stated signal reliability is used more consistently by the reviewer or by a coded update. Reviewer-only, monitor-only, and combined results distinguish complementarity from substitution.

\section{Problem Setting}
\label{sec:problem}

\subsection{Risk and Action Are Different Quantities}

Let $Y=1$ mean that a patch fails the archived evaluation harness. Let $X$ denote the issue, patch, and visible code context. A separate model may also provide evidence $S$, such as a failure score. Finally, let $\pi$ denote the assumed failure rate before examining the patch. The probability reported by a reusable risk estimator is
\begin{equation}
 p_{\pi}(X,S)=\Prob_{\pi}(Y=1\mid X,S).
 \label{eq:risk}
\end{equation}
Changing $X$, $S$, or the assumed failure rate may legitimately change this probability.

Decision costs play a different role. If accepting a failing patch costs $C_{\mathrm{FA}}$ and rejecting a passing patch costs $C_{\mathrm{FR}}$, the stated loss is minimized by rejecting when
\begin{equation}
 p_{\pi}(X,S)\geq \tau(C)
 =\frac{C_{\mathrm{FR}}}{C_{\mathrm{FA}}+C_{\mathrm{FR}}}.
 \label{eq:threshold}
\end{equation}
This is the standard Bayes decision rule for binary actions \citep{berger1985statistical}. Equal costs give $\tau=.50$; a 10:1 false-accept penalty gives $\tau=1/(10+1)\approx .091$. Thus, a cost change should alter the action threshold, not the probability for fixed $X$, $S$, and $\pi$.

We call a reported probability \emph{policy-dependent} when it changes after only the downstream cost-and-threshold block changes. This is not the same as ordinary miscalibration. A miscalibrated score can still have a stable meaning across policies; a policy-dependent score cannot.

\begin{figure*}[t]
\centering
\includegraphics[width=\textwidth]{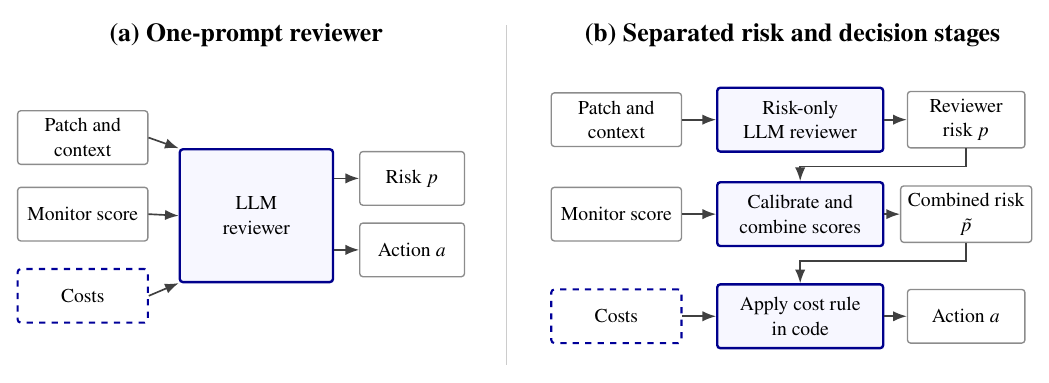}
\caption{System designs evaluated in the paper. The monitor is a second model that assigns a failure score to each patch. In (a), one prompt presents the patch, monitor score, and costs, then requests both a probability and an action. In (b), the reviewer reports risk without policy information; code calibrates and combines the reviewer and monitor scores, then applies the cost rule.}
\label{fig:architecture}
\end{figure*}

Figure~\ref{fig:architecture} shows the two system designs studied in this paper. The one-prompt design asks one model for both a probability and an action after showing it the patch, monitor score, and costs. The modular design requests patch risk without policy information, combines the monitor separately, and applies the cost rule in code. We test the modular design as a sufficient, inspectable alternative; we do not claim that separate calls are the only possible solution.

\section{Experimental Setup}
\label{sec:setup}

\subsection{Data, Reviewers, and Monitor}

We use 360 Python-repository issues from SWE-rebench \citep{badertdinov2025swerebench}. Repositories are disjoint across calibration set A (80 issues, 45 repositories), calibration set B (80 issues, 42 repositories), and the test set (200 issues, 114 repositories). For each issue, deterministic filtering retains one Qwen3-Coder/OpenHands patch that passed and one that failed the archived executable harness \citep{qwen2025coder,wang2025openhands}. The 720 patches were generated before reviewer evaluation. The label is the archived harness outcome, not a complete judgment of security, maintainability, or semantic correctness. Because every issue contributes one pass and one failure, the candidate-level failure rate is 50\% by construction.

We evaluate \DeepSeek, \Grok, \Mistral, and \GPT \citep{deepseek2026v4,xai2026grok43,mistral2026medium35,openai2026gpt55}. The fixed deployments were Azure version 2026-04-23, Azure version 1, Azure GlobalStandard version 1, and Codex CLI 0.145.0 with medium reasoning, respectively. Each reviewer sees the issue, the complete normalized patch, and fixed base-code context around edited regions. Calls are stateless and tool-free. A separate model, Gemini 3.5 Flash with a pinned May 19, 2026 realization, provides one continuous failure score for each patch; we refer to this model as the monitor \citep{google2026gemini35}.

Each reviewer produced 3,948 valid responses: 720 risk-only calls, 2,188 controlled calls, 1,000 operational calls, and 40 repeat measurements. The risk-only prompt asks only for a failure probability and rationale. It shows no monitor score, assumed failure rate, error cost, threshold, or action request.

\subsection{Prompt Conditions}

Table~\ref{tab:conditions} lists the four decision contexts. We use descriptive names in the text; A--D are the identifiers in the release. The low-prior condition changes the assumed failure rate while leaving the action policy unchanged. Because the test set remains balanced, this arm is a directional control rather than a calibration test for a population with 10\% failures. The cost-10 and cost-20 conditions change the false-accept cost and the corresponding printed threshold together.

\begin{table}[t]
\centering
\small
\setlength{\tabcolsep}{3.2pt}
\begin{tabular}{@{}llrrc@{}}
\toprule
ID & Condition & Prior & $C_{\mathrm{FA}}$ & Threshold \\
\midrule
A & Equal cost & .50 & 1 & $1/2$ \\
B & Low prior & .10 & 1 & $1/2$ \\
C & Cost 10 & .50 & 10 & .091 \\
D & Cost 20 & .50 & 20 & .048 \\
\bottomrule
\end{tabular}
\caption{Prompt conditions. The false-reject cost is 1 in every condition. Thresholds are rounded here for readability; the prompts use the exact cost-derived values. Condition B changes the assumed failure rate and is not an intermediate cost setting.}
\label{tab:conditions}
\end{table}

The operational test panel contains 400 matched equal-cost/cost-10 pairs per reviewer. Each pair uses the same issue, patch, and realized monitor card. A controlled panel contains 160 matched equal-cost/low-prior/cost-10 cells per reviewer over 33 repositories. Condition order is balanced in the controlled panel, and operational calls are interleaved in recorded order. Because cost and threshold change together, the main intervention identifies the effect of their joint policy block rather than the separate effect of either line.

Table~\ref{tab:prompt} summarizes the shared endpoint definition, the two policy blocks in the main comparison, and the shared question. Within a matched pair, the issue, patch, code context, monitor card, system message, question, schema, and all remaining user-message bytes are identical. Strict JSON schemas yield one accepted response for every planned condition. Retries do not add scientific observations.

\begin{table*}[t]
\centering
\small
\setlength{\tabcolsep}{5pt}
\begin{tabular}{@{}p{0.47\textwidth}p{0.47\textwidth}@{}}
\toprule
\multicolumn{2}{@{}p{0.96\textwidth}@{}}{\textbf{Shared endpoint instruction.} \texttt{HARNESS\_FAILING means the candidate would fail the frozen evaluation harness; HARNESS\_PASSING means it would pass that harness. Estimate only that event from the evidence supplied.}} \\
\midrule
\textbf{Equal-cost block} & \textbf{Cost-10 block} \\
{\ttfamily\raggedright Assumed deployment base rate:\newline P(HARNESS\_FAILING)=0.500000.\newline Loss for ACCEPT when\newline HARNESS\_FAILING=1.000000.\newline Loss for REJECT when\newline HARNESS\_PASSING=1.000000.\newline Reject when posterior\newline P(HARNESS\_FAILING) $\geq$ 0.500000000000;\newline at equality choose REJECT.}
&
{\ttfamily\raggedright Assumed deployment base rate:\newline P(HARNESS\_FAILING)=0.500000.\newline Loss for ACCEPT when\newline HARNESS\_FAILING=10.000000.\newline Loss for REJECT when\newline HARNESS\_PASSING=1.000000.\newline Reject when posterior\newline P(HARNESS\_FAILING) reaches the cost-derived threshold ($\tau_{10}\approx .091$);\newline at equality choose REJECT.} \\
\midrule
\multicolumn{2}{@{}p{0.96\textwidth}@{}}{\textbf{Shared question.} \texttt{What is the probability from 0 to 1 that this candidate is HARNESS\_FAILING after using all displayed evidence? Choose ACCEPT or REJECT for this candidate under the displayed deployment setting.} The response schema requires \texttt{failure\_probability}, \texttt{decision}, and a bounded \texttt{rationale}.} \\
\bottomrule
\end{tabular}
\caption{Prompt excerpt for the main comparison. The threshold is rounded to three decimals in this table; the experiment used the exact 12-decimal value shown in the rendered prompts. Issue-specific evidence appears between the policy block and the shared question.}
\label{tab:prompt}
\end{table*}

\subsection{Metrics and Statistical Analysis}

We report two main outcomes. The Brier score is the mean squared error of a probability forecast, so lower values indicate more accurate probabilities. Decision loss uses the costs printed in the prompt: accepting a failing patch costs $c$, rejecting a passing patch costs 1, and correct actions cost 0. Loss per issue sums the two retained patches; rejecting both therefore has loss 1. We use \emph{acceptance rate} for the fraction of candidate patches accepted.

Calibration set A selects the cutoff that converts the continuous monitor score into a WARNING/CLEAR card. Calibration set B measures the performance of that fixed card. Continuous score calibration and combination use both calibration sets. When choosing among model forms, each calibration repository is predicted by a model that was not trained on that repository. For raw reviewer score $r$ and monitor score $m$, let $z_R=\operatorname{logit}(\operatorname{clip}r)$ and $z_M=\operatorname{logit}(\operatorname{clip}m)$, with clipping to $[.001,.999]$. The combined model is
\begin{equation}
 \widehat p=\sigma(\beta_0+\beta_R z_R+\beta_M z_M).
 \label{eq:fusion}
\end{equation}
All fitted models are frozen before the test set is evaluated.

Confidence intervals use 20,000 bootstrap samples of repositories, preserving all matched calls from each sampled repository. For a claim that must hold for every reviewer, we use an intersection--union test: the joint claim succeeds only if every reviewer passes its one-sided test. For fitted pipelines, the primary bootstrap keeps the calibration-trained model fixed; a nested check also resamples calibration repositories and refits. Deterministic policies compare the reported decimal with the exact decimal threshold printed in the prompt.

\section{Results}
\label{sec:results}

\begin{figure*}[t]
\centering
\includegraphics[width=\textwidth]{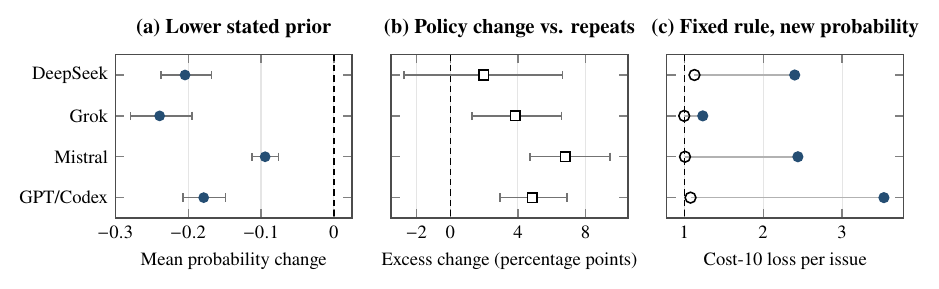}
\caption{Matched effects of prior and policy. Points show means and bars show 95\% repository-bootstrap intervals. (a) Lowering the stated prior from .50 to .10 lowers reported risk. (b) Policy-induced movement minus same-prompt repeat movement; positive values exceed the measured run-to-run baseline. (c) Loss when the same cost-10 rule is applied to cost-10 probabilities (filled) or matched equal-cost probabilities (open). Lower loss is better; the dashed line marks reject-all.}
\label{fig:policy}
\end{figure*}

\begin{table*}[t]
\centering
\small
\setlength{\tabcolsep}{6pt}
\begin{tabular}{@{}lrrrr@{}}
\toprule
& \multicolumn{3}{c}{Loss per issue at 10:1 cost} & \\
\cmidrule(lr){2-4}
Reviewer & Model action & Cost-10 probability & Equal-cost probability & Reduction [95\% CI] \\
\midrule
DeepSeek & 2.550 & 2.400 & 1.130 & 1.270 [0.724, 1.816] \\
Grok & 1.365 & 1.235 & 1.000 & 0.235 [0.037, 0.464] \\
Mistral & 2.480 & 2.440 & 1.010 & 1.430 [0.934, 1.943] \\
GPT/Codex & 3.740 & 3.530 & 1.080 & 2.450 [1.840, 3.072] \\
\bottomrule
\end{tabular}
\caption{Decision loss under a fixed high-cost rule. ``Model action'' is the action returned by the cost-10 prompt. The next two columns apply the same exact cost-10 rule in code to probabilities from the cost-10 and equal-cost prompts. Reduction is the difference between those deterministic losses. Reject-all loss is 1.}
\label{tab:fixed-threshold}
\end{table*}

\begin{figure*}[t]
\centering
\includegraphics[width=\textwidth]{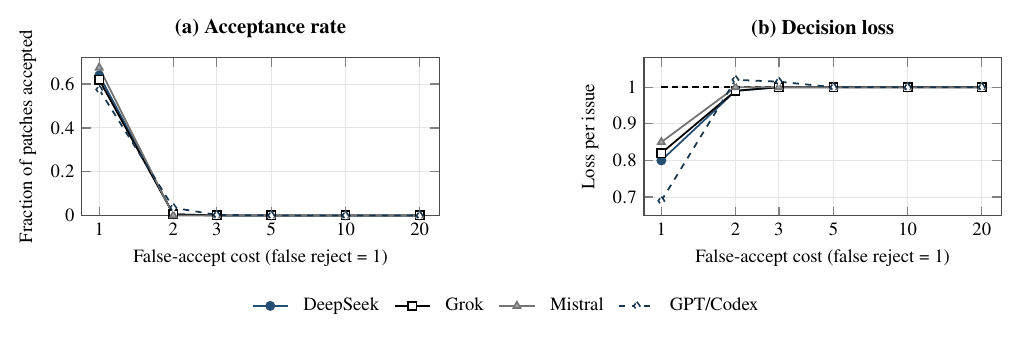}
\caption{Cost--acceptance curve for the modular pipeline. Each line corresponds to one reviewer after external combination with the monitor. Panel (a) shows the fraction of patches accepted; panel (b) shows loss per issue, with reject-all at 1. At equal costs, acceptance is 58--68\% and loss is .69--.85. As false-accept cost rises, acceptance falls to zero and loss approaches reject-all.}
\label{fig:frontier}
\end{figure*}

\begin{figure*}[t]
\centering
\includegraphics[width=\textwidth]{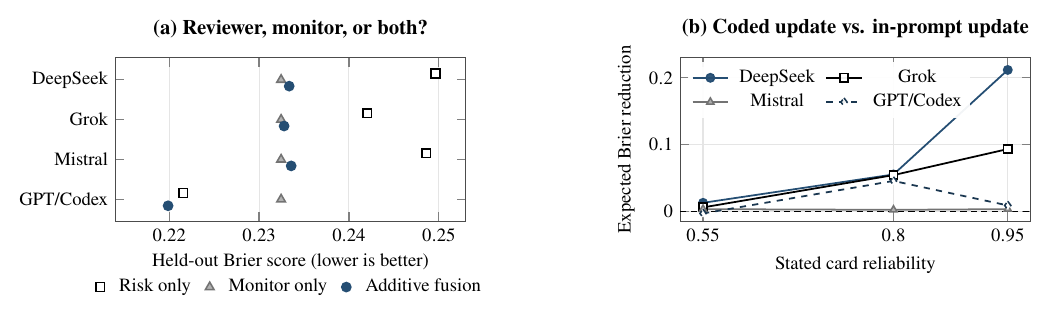}
\caption{Component and evidence-update tests. (a) Held-out Brier scores for the risk-only reviewer, monitor, and additive fusion; lower is better. The monitor is best for DeepSeek, Grok, and Mistral. Fusion is best for GPT/Codex, but the gain is uncertain. (b) Expected Brier reduction from applying the stated card reliability in code rather than in the reviewer prompt; positive values favor code. Coded updates improve 11 of 12 reviewer--reliability cells.}
\label{fig:evidence}
\end{figure*}

\subsection{Policy Prompts Change Reported Risk}

\paragraph{The prior produces the expected response.}
Lowering the assumed failure rate from .50 to .10 lowers mean reported risk by .204 [.168,.237] for DeepSeek, .239 [.195,.279] for Grok, .094 [.076,.112] for Mistral, and .179 [.149,.207] for GPT/Codex (Figure~\ref{fig:policy}a).

\paragraph{The downstream policy also changes the reported probability.}
Replacing the equal-cost block with the joint 10:1 cost and corresponding threshold block changes matched probabilities by .148, .136, .169, and .153 on average. Exact agreement between the two calls is only 31.0\%, 11.5\%, 11.0\%, and 8.5\%. The signed direction differs across reviewers, so higher costs do not induce one common upward or downward bias.

Two calls can differ even when the prompt is unchanged. We therefore compare the policy-induced movement with same-prompt repeats. The excess movement is .020 [$-.027$,.066] for DeepSeek, .038 [.013,.066] for Grok, .068 [.047,.095] for Mistral, and .049 [.029,.069] for GPT/Codex (Figure~\ref{fig:policy}b). The policy effect is larger than measured repeat variation for three systems and unresolved for DeepSeek. Averaged across reviewers, the cost-10 prompt worsens Brier score by .0323 [.0207,.0437].

Mistral provides a concrete example of how the policy can enter the reported number. Under CLEAR evidence, it reports the exact cost-10 threshold in 37 of 80 cases when that value is printed and in 0 of 80 matched controls. When the prompt switches to the cost-20 threshold, exact reports rise from 1 of 40 to 18 of 40. Every exact-boundary report is paired with \textsc{reject}, as instructed at equality. The experiment cannot separate copying, anchoring, and action-driven reporting.

\subsection{Probability Source Drives High-Cost Loss}

The decisions returned by the cost-10 prompts have loss 2.550, 1.365, 2.480, and 3.740 per issue. All four exceed the reject-all loss of 1 (intersection--union $p=.0014$). Grok, Mistral, and GPT/Codex nevertheless agree with deterministic thresholding of their reported probabilities on 98.25\%, 88.25\%, and 96.75\% of calls. The problem is therefore not only failure to apply the stated action rule.

We next hold the rule itself fixed. Code applies the exact cost-10 threshold shown in the prompt to either the probability from the cost-10 prompt or the matched probability from the equal-cost prompt. The threshold and its implementation are identical; only the prompt that produced the probability differs. The equal-cost probability yields lower loss for every reviewer (Table~\ref{tab:fixed-threshold}; Figure~\ref{fig:policy}c). Each reviewer-level interval excludes zero, and the all-reviewer intersection--union test gives $p=.0063$.

This comparison identifies excess loss associated with showing the high-cost block while eliciting the probability. It does not require either probability to be calibrated. It also does not provide a useful high-cost approval policy: at this threshold, the equal-cost probabilities accept at most 2.25\% of candidates.

\subsection{Modular Composition Improves Equal-Cost Performance}

\paragraph{Equal-cost performance improves, but high-cost coverage vanishes.}
We evaluate the design in Figure~\ref{fig:architecture}b directly. The reviewer call requests only patch-failure probability. We calibrate that score, optionally combine it with the monitor, and apply the cost-derived threshold in code. At equal costs, the combined pipeline accepts 64.0\%, 62.0\%, 67.5\%, and 57.5\% of candidates, with loss .800, .820, .850, and .690 per issue. Averaged across reviewers, its loss is .0725 [.0285,.1167] below calibrated reviewer-only scoring. A nested bootstrap that refits the calibration models gives .0729 [.0249,.1291].

Acceptance falls quickly as false-accept cost rises (Figure~\ref{fig:frontier}). At cost 10, reviewer-only, monitor-only, binary-card, and continuous-score pipelines all reject every patch and match reject-all loss of 1. The modular pipeline avoids the large excess losses of the one-prompt system, but the available scores are not accurate enough to support nontrivial approval at this cost.

\paragraph{The monitor often substitutes for a weaker reviewer.}
Figure~\ref{fig:evidence}a compares the risk-only reviewer, monitor-only, and fused scores. The monitor has the lowest Brier point estimate for DeepSeek, Grok, and Mistral. For GPT/Codex, the reviewer is better than the monitor; fusion lowers Brier from .2216 to .2199, but the interval for this small gain includes zero. The interaction term changes Brier score by less than .0003 in every system; all intervals cross zero.

The continuous-score comparison is not a like-for-like test: the modular pipeline receives the full monitor score, whereas the joint prompt sees only a WARNING/CLEAR card. We therefore run a second experiment in which both methods receive the same binary card and the same stated reliability $q\in\{.55,.80,.95\}$. One method asks the reviewer to revise its probability in the prompt. The other applies the odds multiplier implied by the card's stated accuracy:
\begin{equation}
 \mathrm{LR}_{\mathrm{WARN}}=\frac{q}{1-q},\qquad
 \mathrm{LR}_{\mathrm{CLEAR}}=\frac{1-q}{q}.
 \label{eq:lr}
\end{equation}
The calculated update improves expected Brier score in 11 of 12 reviewer-by-reliability cells (Figure~\ref{fig:evidence}b); the only reversal is GPT/Codex at $q=.55$. Averaged across reviewers, expected Brier improves by .0411 [.0271,.0552] and expected log loss by .1146 [.0740,.1559]. The four reviewers respond differently to the same WARNING/CLEAR evidence, so the data do not support one common warning bias.

\section{Robustness Checks}
\label{sec:robustness}

We use several checks to distinguish policy dependence from simpler explanations (Table~\ref{tab:alternatives}).

\begin{table*}[t]
\centering
\small
\setlength{\tabcolsep}{5pt}
\begin{tabular}{@{}p{0.20\textwidth}p{0.28\textwidth}p{0.44\textwidth}@{}}
\toprule
Potential explanation & Test & Finding \\
\midrule
Call-to-call variation & Compare the policy change with unchanged-prompt repeats & Excess movement is clear for Grok, Mistral, and GPT/Codex; DeepSeek remains unresolved on this comparison. \\
Failure to apply the action rule & Compare model-returned actions with deterministic thresholding, then hold the rule fixed & Three systems follow the rule on 88--98\% of calls, yet their loss remains high; changing only the probability source lowers loss for all four. \\
Reviewer capability & Compare policy-free probability quality with model-returned high-cost loss & GPT/Codex has the best reviewer-only Brier score but the worst cost-10 loss. Reviewer capability and interface quality are different properties. \\
Monitor complementarity & Compare reviewer-only, monitor-only, and fused probabilities & The monitor largely replaces the three weaker reviewers and provides only a small, uncertain gain over GPT/Codex alone. \\
Simple fusion is too weak & Compare additive fusion with interaction, nonlinear, and routing alternatives chosen on calibration data & The product term changes Brier score by less than .0003 in every system; no more complex alternative improves reliably over the strongest simple model. \\
\bottomrule
\end{tabular}
\caption{Alternative explanations and the tests used to examine them. The repeat control, fixed-rule comparison, component baselines, and model-family checks address different threats to the interpretation; remaining scope limits appear in Section~\ref{sec:limitations}.}
\label{tab:alternatives}
\end{table*}

The fitted-pipeline result is also stable to calibration-sample uncertainty. A nested bootstrap resamples calibration repositories, refits the models, and independently resamples test repositories. It retains an average Brier improvement of .0100 [.0039,.0175] and an equal-cost loss improvement of .0729 [.0249,.1291]. The modular result is therefore not tied to one favorable calibration split. The individual GPT/Codex gain remains small and uncertain, so we describe the monitor as a weak complement for that reviewer rather than a uniform improvement.

Reviewer behavior also differs in mechanism and direction. Three systems lower their reported probability on average under the high-cost block, while Grok moves slightly upward. Mistral frequently reproduces the printed boundaries, whereas Grok and GPT/Codex do not. DeepSeek and Mistral respond to the same WARNING/CLEAR evidence in opposite directions. The shared result is policy dependence, not one common conservative shift or warning bias.

\section{Discussion}

\subsection{Implications for System Design}

The results separate an interface problem from a capability limit. The interface problem is that a reported probability changes when the downstream policy changes. The capability limit remains after that coupling is removed: even a stable score may not separate passing and failing patches well enough to support approvals when false acceptance is very costly. The fixed-rule comparison measures the first problem; the cost--acceptance curve exposes the second.

Reviewer quality is not one number. GPT/Codex gives the best policy-free reviewer score but the worst model-returned cost-10 loss. Grok and GPT/Codex usually convert their own probabilities into actions consistently with the displayed threshold, yet those decisions remain poor. The monitor has a lower Brier point estimate than three reviewers, and adding those reviewers does not improve the point estimate over the monitor alone. Risk ranking, probability accuracy, evidence complementarity, and decision utility are separate properties; a leaderboard based on one can hide a failure in another.

The modular design assigns a testable role to each output. The reviewer reports risk without seeing the policy, the monitor contributes a separate measurement, and the controller applies a declared loss function. This makes the score reusable when costs change, reveals the monitor's marginal value, and turns the final action into deterministic code. It also exposes negative results that an end-to-end score can hide: the monitor mostly replaces weaker reviewers, more complex fusion shows no reliable gain, and at a 10:1 penalty every tested controller has zero acceptance.

Separate calls are not the only possible implementation. A joint prompt might work if it defines probability independently of action, commits to risk before showing costs, or returns fields that are scored separately. Such a design should still pass the same policy-change and component tests. The practical requirement is a stable risk estimate, not a particular number of model calls.

A policy-dependent score is difficult to use as a software interface. It may be cached, recalibrated, compared across versions, or consumed by teams with different costs. If the elicitation policy is already folded into the number, changing the controller can silently change what the score means. Separating the stages leaves a clearer record of whether a change came from the reviewer, the monitor, or the decision policy.

\subsection{How to Evaluate Reviewer Systems}

Three reports are especially useful. First, hold the case and evidence fixed, vary the downstream policy, and compare the resulting probability movement with repeated calls under an unchanged prompt. Second, report reviewer-only, monitor-only, and combined probability error; when internal and external evidence integration are compared, both methods should receive the same representation. Third, apply the decision rule outside the model and report both acceptance and loss over the cost range relevant to deployment. Together, these checks distinguish an unstable probability, a redundant monitor, and a score that is too weak for the desired operating point.

The paired benchmark is well suited to the matched policy and fixed-rule comparisons because each issue contributes one passing and one failing patch under the same repository context. It is less suited to estimating production prevalence or the absolute value of an approval policy. The within-patch comparisons do not require a natural failure rate, whereas calibration and coverage at a particular cost do. The operating curve should therefore be read as the capability of these fitted scores on a balanced case mix, not as a production forecast.

The same evaluation question applies beyond code review whenever an LLM-generated probability is passed to a system with its own costs, such as safety judging, triage, fraud screening, reward modeling, or deferral. Our data do not establish behavior in those domains. They provide a direct test to run there: when the evidence is unchanged, does the reported probability remain a statement about the case, or does it move with the policy that consumes it?

\section{Limitations}
\label{sec:limitations}

The cost and printed threshold vary together. The study therefore identifies the joint policy block, not the separate effects of cost semantics, threshold salience, or choosing an action first. A factorial experiment that varies cost text, threshold text, and irrelevant numbers would distinguish these explanations.

The benchmark uses an archived harness label and one passing plus one failing patch per issue. It is not a natural sample of pull requests, so its absolute calibration and reject-all comparisons do not transfer directly to production. Reviewers also receive static context without tools. Deployment studies should include natural failure rates, both-pass and both-fail cases, no-valid-patch cases, tool use, latency, and monetary cost.

We evaluate one generator stream, one monitor, and four fixed reviewer deployments; the results do not establish a population-wide law. At a 10:1 penalty, the modular pipeline accepts no patches, and the repeat panel is too small to resolve DeepSeek's effect beyond run-to-run variation.

\section{Conclusion}

Across four reviewer interfaces, changing the joint cost-and-threshold block changed the failure probability reported for the same patch and evidence. The actions returned by the cost-10 prompt were worse than reject-all for every reviewer. Under the same high-cost rule, matched equal-cost probabilities reduced loss for all four systems, showing that probability elicitation contributes to the excess loss. Relative to calibrated reviewer-only scores, the policy-free pipeline improved average probability accuracy and equal-cost decision loss. The monitor mostly replaced weaker reviewers; at 10:1, the pipeline accepted no patches and matched reject-all. These results support evaluating the risk estimate, outside evidence, and final action separately.

\section*{Acknowledgments}

This work was conducted through the Algoverse AI Research Program. Funding from the program supported the computational costs of the study.

\appendix

\section{Additional Data and System Details}
\label{sec:supp-data}

\subsection{Tasks and Splits}

The corpus contains 360 Python-repository issues derived from \texttt{nebius/SWE-rebench} \citep{badertdinov2025swerebench} at pinned revision \texttt{89cdfbab4ab1}; the full hash is recorded in the public repository. The issues span 201 repositories. A repository contributes at most five issues and appears in exactly one split: CAL\_A contains 80 issues from 45 repositories, CAL\_B contains 80 issues from 42 repositories, and EVAL contains 200 issues from 114 repositories. The source cutoff was 17 July 2026 UTC. Included repositories use MIT, Apache-2.0, BSD-3-Clause, or BSD-2-Clause licenses.

\subsection{Patch Selection and Outcome}

Candidate patches come from the public OpenHands trajectory stream \path{openhands_0_54_0_qwen3_coder_480b_a35b} \citep{qwen2025coder,wang2025openhands}, pinned at revision \texttt{35455389ab51}. Deterministic filters retain one archived harness pass and one archived harness failure per issue after checking patch structure, applicability, duplication, similarity to the gold patch, hidden-test paths, sensitive content, and prompt-size limits. Candidate position is balanced: each position contains 180 passes and 180 failures. Neither the gold patch, test patch, nor archived outcome is shown to the monitor or reviewers.

The outcome is the upstream executable-harness \texttt{resolved} field: 1 maps to \textsc{Harness-Pass} and 0 to \textsc{Harness-Fail}. Corpus construction reapplies every patch and verifies structural eligibility, but it does not rerun the full benchmark harness for all 720 candidates. The outcome is therefore an archived executable result, not a complete judgment of semantic correctness, security, maintainability, or production safety.

\subsection{What Reviewers See}

For every candidate, the monitor and reviewers receive the issue statement, the complete normalized diff, and deterministic base-code context with a 40-line radius around changed regions. They have no repository shell, tests, browser, hidden files, external tools, or memory across calls. The generator used agentic tools; the monitor and reviewers evaluate a fixed rendering.

\input{tables/table_systems.tex}

\section{Prompt Conditions and Call Inventory}
\label{sec:supp-design}

\subsection{Call Inventory and Response Formats}

\input{tables/table_full_design.tex}
\input{tables/table_interfaces.tex}
\input{tables/table_arms.tex}

The risk-only calls ask only for the probability of archived harness failure from the issue, patch, and visible code context. They show no monitor card, stated prior, error cost, threshold, or action request. The response contains \texttt{failure\_probability} and a bounded rationale. There are 720 risk-only calls per reviewer: 160 on CAL\_A, 160 on CAL\_B, 400 on EVAL, and 40 repeats. These calls provide the reviewer score used in the calibration and monitor analyses.

The equal-cost probability used in the fixed-threshold comparison is different. It comes from a joint prompt that asks for both probability and action while showing prior .50, equal error costs, and threshold $1/2$. It is a matched diagnostic, not the risk-only estimator.

\subsection{Prompt Conditions}

The release uses A--D as identifiers for four conditions. A is the equal-cost condition (prior .50 and equal error costs). B is the low-prior control (prior .10 and equal error costs). C and D increase the false-accept cost to 10 and 20 and print the corresponding cost-derived thresholds. The exact 12-decimal strings are preserved in the rendered prompts and the threshold-precision audit. The letters are identifiers, not an ordered cost scale. CAL\_B is the second calibration split and is unrelated to condition B.

\subsection{Controlled Evidence Cards}

The controlled cards have symmetric reliability $q\in\{.55,.80,.95\}$:
\begin{equation}
 \Prob(\textsc{Warning}\mid Y=1)=q,\qquad
 \Prob(\textsc{Clear}\mid Y=0)=q.
 \label{eq:supp-card-generator}
\end{equation}
The corresponding likelihood ratios are
\begin{equation}
 \mathrm{LR}_{\textsc{Warning}}=\frac{q}{1-q},\qquad
 \mathrm{LR}_{\textsc{Clear}}=\frac{1-q}{q}.
 \label{eq:supp-card-lr}
\end{equation}
Additional arms vary lexical labels, table versus prose rendering, asymmetric sensitivity and specificity, output order, base rate, and false-accept cost. These cards are controlled interventions; they do not describe the empirical accuracy of the operational monitor.

\subsection{Operational Monitor Card}

The operational monitor supplies one continuous score for each of the 720 patches; the release identifier for this score is M1. A WARNING/CLEAR cutoff is selected on CAL\_A by maximizing Youden's $J$; ties favor higher specificity and then the larger numerical threshold. The selected rule is WARNING when M1 risk is at least .10. Sensitivity and specificity are .513 and .725 on CAL\_A, .488 and .562 on CAL\_B, and .545 and .650 on EVAL. The external model keeps the continuous score. Joint operational prompts receive the binary rendering.

\section{Prompt Execution and Parsing}
\label{sec:supp-execution}

The main text reproduces the invariant system instruction, both policy blocks, the shared question, and the response fields. The complete issue-specific prompts are included in the public repository. In the controlled A--C prompt set, the system message is identical in all 640 reviewer--candidate--signal pairs. After replacing the two cost-and-threshold lines, every remaining user-message byte is identical. The repository also contains 320 A--D pairs and 320 C--D pairs with the same property.

\input{tables/table_prompt_audit.tex}

The A/B/C control panel contains 160 matched candidate--signal cells per reviewer over 33 repositories. Each condition appears first in 80 cells and second in 80 cells for every reviewer. Order is fixed independently of reviewer responses. Operational equal-cost and cost-10 calls are interleaved in recorded stage order: 407--413 condition runs occur across 800 calls per reviewer, and the longest run of one condition is 8--10 calls. Comparable wall-clock timestamps are not available across all serving systems, so we do not estimate cross-provider temporal drift.

The analysis dataset contains one accepted response for every planned prompt condition: 3,948 per reviewer, with no missing or duplicate cells. Retries caused by service or formatting events do not create additional observations. GPT/Codex provenance records 3,950 physical attempts for 3,948 accepted calls. Comparable attempt ledgers are unavailable for the other three systems, so retry rates are not compared across reviewers. Strict schemas constrain field type, range, and order. The analysis uses parsed fields rather than rationale text.

The four systems share the logical matrix, rendered evidence, questions, and response schemas. They do not share a serving API or a common set of exposed sampling controls. We therefore report effects of deployed reviewer systems rather than isolated base-model effects. No unrecorded temperature, top-$p$, seed, or reasoning-effort value is imputed across providers.

\section{Metrics and Statistical Inference}
\label{sec:supp-methods}

\subsection{Probability and Decision Outcomes}

Let $Y=1$ denote archived harness failure, $X$ the issue, patch, and visible code context, $S$ monitor evidence, and $\pi$ the assumed failure prevalence. A reusable risk estimate is
\begin{equation}
 p_{\pi}(X,S)=\Prob_{\pi}(Y=1\mid X,S).
 \label{eq:supp-risk-contract}
\end{equation}
Under prior-probability shift, $\pi$ may change the posterior while the class-conditional evidence process remains fixed. For fixed $X$, $S$, and $\pi$, error costs should change the action threshold rather than the reported probability.

For candidate $i$ in repository $g$, let $p^A_{ig}$ and $p^C_{ig}$ denote reports under equal costs and cost 10. We report
\begin{align}
\Delta_{\mathrm{abs}} &= \mathbb{E}|p^C-p^A|,\\
\Delta_{\mathrm{signed}} &= \mathbb{E}[p^C-p^A],\\
\Delta_{\mathrm{Brier}} &= \mathbb{E}[(p^C-Y)^2-(p^A-Y)^2].
\end{align}
Two independent calls can differ even when the prompt is unchanged. The repeat-controlled estimate therefore subtracts same-prompt absolute movement from matched A--C movement.

For false-accept cost $c$ and action $a\in\{\textsc{Accept},\textsc{Reject}\}$,
\begin{align}
\ell_c(a,Y)={}&c\,\mathbb{1}[a=\textsc{Accept},Y=1]\\
&+\mathbb{1}[a=\textsc{Reject},Y=0].
\end{align}
Loss per issue sums the two retained patches. Since each issue has one pass and one failure, reject-all has loss 1. ``Model-returned action'' uses the action emitted in the cost-10 call. ``Cost-10 probability'' applies the exact cost-10 threshold printed in the prompt to the cost-10 probability. ``Equal-cost probability'' applies that same threshold to the matched equal-cost probability.

\subsection{Calibration and Monitor Combination}

Raw scores are clipped to $[.001,.999]$ before logit transformation. Let $r$ be the risk-only reviewer score and $m$ the continuous M1 score:
\begin{equation}
 z_R=\operatorname{logit}(r),\qquad z_M=\operatorname{logit}(m).
\end{equation}
Reviewer-only and monitor-only calibration use one-feature logistic models. The combined model is
\begin{equation}
 \widehat p_{\mathrm{add}}=\sigma(\beta_0+\beta_R z_R+\beta_M z_M).
 \label{eq:supp-additive}
\end{equation}
An interaction sensitivity check adds $\beta_{RM}z_Rz_M$. Binary-card combination replaces $z_M$ with a WARNING/CLEAR feature. CAL\_A selects the operational binary cutoff; CAL\_B measures the performance of that frozen card. Continuous calibration and combination use pooled CAL\_A and CAL\_B, with deterministic repository-fold out-of-fold selection when a model family or calibration form is chosen. Selected models are refit on pooled CAL and frozen before EVAL.

\subsection{Uncertainty Estimates}

Primary confidence intervals use 20,000 repository-cluster bootstrap draws. Each draw retains every issue, candidate, and matched call from a sampled repository. Claims requiring a positive effect in all four reviewers use an intersection--union test; its $p$-value is the largest of the four one-sided reviewer-specific values. Brier score is not clipped. Log loss clips probabilities to $[.001,.999]$. Pair accuracy gives 0.5 credit to exact ties.

For fitted pipelines, the primary bootstrap resamples EVAL repositories while holding the CAL-trained model fixed. A nested sensitivity independently resamples CAL repositories, refits the models, and resamples EVAL repositories.

\section{Policy Prompts and Reported Risk}
\label{sec:supp-rq1}

\subsection{Prior Control and Policy Dependence}

\input{tables/table_prior_control.tex}

The low-prior condition checks whether reviewers respond to information that should affect risk; it is not an intermediate cost. Moving the stated failure prior from .50 to .10 while holding equal costs and threshold $1/2$ lowers reported risk in every system. Because the retained sample is still balanced, this is a directional prior-sensitivity result rather than calibration under a 10\% deployment prevalence.

\input{tables/table_operational.tex}

Mean absolute equal-cost--cost-10 movement is .148, .136, .169, and .153. Exact agreement is 31.0\%, 11.5\%, 11.0\%, and 8.5\%. Signed responses differ: DeepSeek, Mistral, and GPT/Codex move downward, while Grok moves slightly upward. The result is therefore not a common conservative bias. Averaged across reviewers, Brier score worsens by .0323 [.0207,.0437]. Model-returned cost-10 loss is 2.550, 1.365, 2.480, and 3.740; every value is above reject-all, with all-system intersection--union $p=.0014$.

\input{tables/table_repeatability.tex}

The repeat panel contains 40 matched candidate--signal cells per reviewer. Policy-induced movement exceeds same-prompt movement for Grok, Mistral, and GPT/Codex; DeepSeek remains unresolved on this comparison. This qualification applies only to the repeat-controlled movement estimate. DeepSeek still shows observed movement and a positive fixed-threshold loss reduction.

\begin{figure}[t]
\centering
\includegraphics[width=\textwidth]{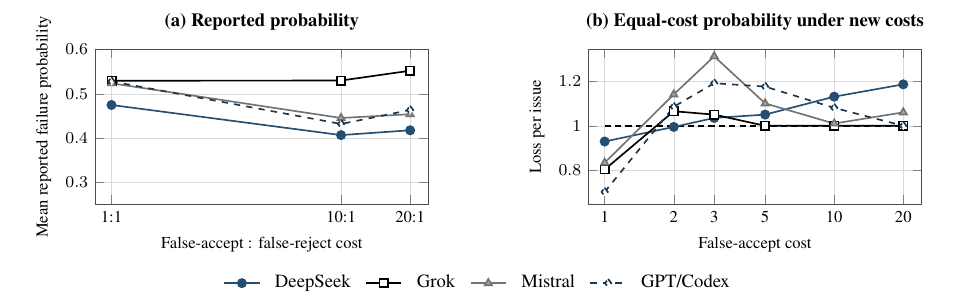}
\caption{Cost-dose response in the controlled panel. Points show reviewer-level mean reported probability under each cost condition; intervals are repository-bootstrap 95\% confidence intervals.}
\label{fig:supp-cost-dose}
\end{figure}

The A--C--D panel does not show a common monotone signed response. Mistral has pronounced threshold-specific behavior; the other systems differ in direction and magnitude. Pooling the signed changes would hide this heterogeneity.

\subsection{Exact Reports of the Printed Threshold}

\input{tables/table_thresholds.tex}

Mistral under CLEAR evidence provides the clearest behavioral pattern. When the cost-10 boundary is printed, exact reports of that boundary rise from 0/80 matched controls to 37/80. When the prompt switches to the cost-20 boundary, exact reports rise from 1/40 in the matched cost-10 condition to 18/40. The paired increases are 46.25 percentage points [35.4,56.8] and 42.5 points [27.3,60.5]. Every exact-boundary report under costs 10 and 20 is paired with \textsc{Reject}, as instructed at equality.

Grok and GPT/Codex never reproduce either unusual boundary in these CLEAR panels, and DeepSeek does so only occasionally. We use ``threshold copying'' as a description of the observed output. The experiment cannot distinguish copying, numerical anchoring, or choosing an action first and then producing a compatible probability. Cost and threshold vary together, so the intervention identifies the joint policy block.

\section{Probability Source and Decision Loss}
\label{sec:supp-rq2}

\subsection{Fixed-Threshold Comparison}

\input{tables/table_stage_inference.tex}

Loss from thresholding the cost-10 probability exceeds loss from thresholding the matched equal-cost probability for all four reviewers; the all-system intersection--union test gives $p=.0063$. Model-returned-action loss minus equal-cost-probability loss is also positive for all four reviewers, but that contrast includes differences in action implementation. At cost 10, the matched equal-cost probability yields losses of 1.130, 1.000, 1.010, and 1.080 while accepting 1.75\%, 0\%, 2.25\%, and 1.50\% of candidates. The comparison isolates one source of excess loss; it does not provide a useful high-cost approval policy.

\subsection{Exact Decimal Thresholds}

The deterministic comparison uses the exact decimal boundary printed in the prompt. This matters for systems that reproduce the boundary. Table~\ref{tab:supp-precision} verifies that every reported policy result uses the canonical decimal string rather than a binary floating-point approximation.

\input{tables/table_precision_audit.tex}

\section{Modular Pipeline and Component Performance}
\label{sec:supp-rq3}

\subsection{Risk-Only Pipeline}

The pipeline uses the risk-only reviewer score and the independent monitor measurement M1. Scores are calibrated or combined on the calibration sets and then frozen before evaluation. The binary-card model receives the same WARNING/CLEAR representation shown in the joint operational prompt; the continuous model keeps the full monitor score.

\input{tables/table_modular_controller.tex}
Figure~\ref{fig:frontier} in the main text reports the same operating frontier; the tables below provide the full numerical values.

At equal costs, continuous combination accepts 57.5--67.5\% of candidates and gives loss .690--.850 per issue. Averaged across reviewers, loss is .0725 [.0285,.1167] below calibrated risk-only scoring, conditional on the fitted CAL pipeline. At cost 10, risk-only, monitor-only, binary-card, and continuous-score pipelines reject every candidate and match reject-all loss of 1.000. The modular pipeline avoids the one-prompt system's excess loss at cost 10, but the available scores are too weak to retain any acceptance at that threshold.

\input{tables/table_nested_refit.tex}

The nested analysis uses 1,000 draws that resample CAL repositories, refit the risk-only and combined models, and independently resample EVAL repositories. It preserves a positive average Brier gain of .0100 [.0039,.0175] and a positive equal-cost loss gain of .0729 [.0249,.1291]. GPT/Codex's individual Brier gain remains small and uncertain. These intervals include calibration-sample uncertainty; the primary intervals describe the frozen CAL-trained pipeline.

\subsection{Reviewer, Monitor, or Both?}

\input{tables/table_monitor.tex}

The monitor alone has the lowest Brier point estimate for DeepSeek, Grok, and Mistral; adding the reviewer does not improve that point estimate. For GPT/Codex, the risk-only reviewer score is stronger than the monitor, and the combination is slightly better than the reviewer alone. In this dataset, the monitor mainly substitutes for weaker reviewers and weakly complements the strongest.

Relative to reviewer-only calibration, the combined model reduces EVAL Brier by .0163 [.0086,.0244], .0093 [.0038,.0149], .0150 [.0078,.0226], and .0017 [$-.0010$,.0045]. The average fitted-pipeline gain is .0106 [.0053,.0161]. Adding $z_Rz_M$ changes Brier by less than .0003 for every reviewer, and every interval crosses zero. Boosting, monotonic boosting, monitor invocation, and reviewer-routing variants show no reliable held-out gain over the strongest simple model.

The continuous comparison gives the external model more information than the joint prompt, which sees only WARNING/CLEAR. It measures the value of preserving and combining the continuous score; it is not a representation-matched test of external versus in-prompt updating.

\subsection{Matched Reliability-Card Experiment}

For the card generator in Equation~\ref{eq:supp-card-generator}, the in-prompt method asks the reviewer to update its probability. The coded method applies
\begin{equation}
 \operatorname{logit}(\widehat p_{\mathrm{updated}})
 =\operatorname{logit}(\widehat p_{\mathrm{code}})+\log\mathrm{LR}(S;q)
 \label{eq:supp-bayes-update}
\end{equation}
outside the LLM. Both possible card realizations are evaluated and weighted by their probabilities under the stipulated generator. The reported values are expected proper scores rather than the result of selecting one favorable card realization.

\input{tables/table_reliability_cells.tex}
\begin{figure}[t]
\centering
\includegraphics[width=\textwidth]{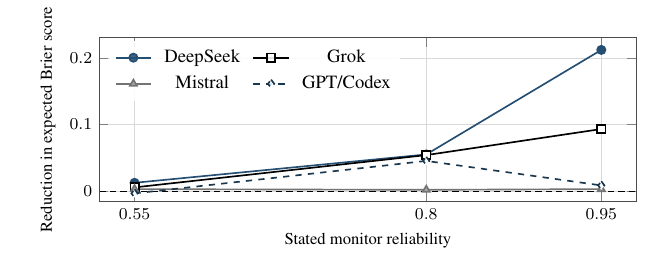}
\caption{Expected Brier-score change from replacing native in-prompt evidence updating with the coded likelihood-ratio update. Positive values favor the coded update; intervals are repository-bootstrap 95\% confidence intervals.}
\label{fig:supp-external-bayes}
\end{figure}

The coded update improves 11 of 12 reviewer--reliability cells. The only reversal is GPT/Codex at $q=.55$. Averaged across reviewers, expected Brier improves by .0411 [.0271,.0552], and expected log loss improves by .1146 [.0740,.1559]. In-prompt WARNING-minus-CLEAR responses differ sharply across systems: the normalized effect is $-1.10$ [$-1.74$,$-.40$] for DeepSeek, $-.10$ [$-.42$,.24] for Grok, $+1.50$ [.72,2.28] for Mistral, and $+.18$ [$-.10$,.47] for GPT/Codex. A pooled near-zero average would be cancellation, not a shared update rule.

\section{Secondary Best-of-Two Analysis}
\label{sec:supp-pairwise}

\input{tables/table_pairwise.tex}

The cascade scores both candidates independently and calls a direct comparator selected on calibration data only when the two scores are exactly equal. The trigger uses no EVAL label. Exact ties occur on 9--31\% of issues. The held-out gain is positive for every evaluated system, and each marginal repository-bootstrap interval excludes zero. In synchronized repository draws, all four gains are positive in 94.6\% of replicates; we therefore report reviewer-level intervals rather than a universal claim.

For DeepSeek, Grok, and Mistral, the calibration data select GPT/Codex as the direct comparator. Always invoking that comparator gives 73.0\% accuracy with a second call on every issue. The cascade reaches 57.5\%, 66.0\%, and 59.0\% with escalation on 10.5\%, 10.0\%, and 31.0\%. GPT/Codex uses itself under a direct-comparison prompt: always-direct accuracy is 73.0\%, while the self-cascade reaches 76.0\% with 9.0\% escalation. Exact ties may reflect score quantization rather than equal latent uncertainty. This result applies to best-of-two or best-of-$N$ selection, not to approval of a single patch or candidate sets in which every patch fails.

\section{Robustness Checks and Negative Results}
\label{sec:supp-negative}

\subsection{No Common Warning Bias}

The four systems do not share a common WARNING-over-CLEAR response. DeepSeek and Mistral have clear effects in opposite directions; Grok and GPT/Codex are near zero. The data therefore do not support a wrapper designed to remove one pooled directional warning bias.

\subsection{No Reliable Gain from More Complex Models}

Across reviewer subsets, raw and logit pooling, medians, votes, calibrated averaging, logistic monitor models, gradient boosting, monotonic boosting, monitor invocation, and reviewer routing, the nontrivial ensemble selected on calibration data ties the strongest single reviewer at 73.5\% pair accuracy. The best method-specific point estimate reaches 74.5\%, a one-point gain whose 95\% interval spans zero. These sweeps set a boundary on the evidence; they are not presented as confirmatory improvements.

\subsection{Limits of the Harness Label}

A qualitative audit identifies cases in which the archived harness is narrower than an intuitive semantic judgment. One issue permits an alternative rejected by the archived harness; another pair has byte-identical production diffs but different auxiliary files. We therefore describe the positive class as an ``archived harness-passing candidate'' rather than an exhaustive semantic oracle.

\section{Reproducibility and Scope}
\label{sec:supp-reproduction}

The public repository contains 15,792 canonical reviewer rows, 720 frozen monitor scores (M1 in the release), task and candidate metadata, repository-disjoint split files, 3,548 materialized prompt templates, the 12,564-row logical-call plan, response schemas, scripts, expected results, dependency specifications, and integrity manifests. Running \texttt{./run\_all.sh} reconstructs the article-facing analyses in a new output directory. Running \texttt{python verify\_release.py} checks corpus structure, exact-decimal threshold decisions, central numerical results, risk-only pipeline frontiers, monitor analyses, candidate selection, and every manifest hash.

The team-authored analysis code is released under the MIT License. Original model outputs, annotations, and derived research tables are released under CC BY 4.0. Third-party benchmark records, repository code, and trajectory material remain under their upstream licenses and are not relicensed. The repository contains the applicable license texts and a third-party notice.

All author-controlled analysis and manuscript regeneration was independently verified in a CPU-only x86\_64 environment with five AMD EPYC 9V74 virtual CPUs, 6~GB RAM, Debian GNU/Linux 13, and Python 3.13.5. No GPU is required for the released analyses. Model inference used the managed deployments listed in Table~\ref{tab:supp-systems}; provider-side physical hardware is not exposed to callers and is not an experimental parameter. The repository records all available deployment identifiers, software versions, seeds, parameter grids, selection criteria, and final settings.

The measured objects are four named reviewer systems, one monitor, one generator stream, and a filtered repository benchmark. Across-reviewer averages are descriptive; the four systems are not a random sample from a model population. Public task and trajectory sources also make unknown training exposure possible. Static no-tool evaluation differs from interactive code review. Deployment studies should include varied prevalence, both-pass and both-fail candidate sets, no-acceptable-candidate cases, dynamic repository access, latency, monetary cost, and a new factorial experiment that separates semantic cost, displayed threshold, and action timing.

Within those limits, the data support three conclusions. First, the joint policy block changes reported failure probabilities while patch evidence is fixed. Second, under the same deterministic threshold, the matched equal-cost probability yields lower loss than the cost-10 probability in all four systems. Third, combining a policy-free reviewer score with the monitor improves average probability accuracy and equal-cost decision loss relative to reviewer-only scoring, while at a 10:1 false-accept cost every tested modular controller rejects all patches and matches reject-all.

\FloatBarrier
\bibliographystyle{plainnat}
\bibliography{references}
\end{document}

%% file: tables/table_systems.tex
\begin{table*}[t]
\centering
\small
\setlength{\tabcolsep}{5pt}
\begin{tabular}{@{}lll@{}}
\toprule
Role & System & Recorded execution surface \\
\midrule
Generator & Qwen3-Coder-480B-A35B-Instruct \citep{qwen2025coder} & OpenHands 0.54.0; vLLM 0.9.0 \\
Operational monitor & Gemini 3.5 Flash \citep{google2026gemini35} & Google AI Studio; pinned realization 20260519 \\
Reviewer 1 & DeepSeek-V4-Pro \citep{deepseek2026v4} & Azure; version 2026-04-23 \\
Reviewer 2 & Grok-4.3 \citep{xai2026grok43} & Azure; version 1 \\
Reviewer 3 & Mistral-medium-3.5 \citep{mistral2026medium35} & Azure GlobalStandard; version 1 \\
Reviewer 4 & GPT-5.5/Codex \citep{openai2026gpt55} & Codex CLI 0.145.0; medium reasoning \\
\bottomrule
\end{tabular}
\caption{Systems and execution surfaces. The generator precedes the study and is not a reviewer. All reviewer calls use static context and no tools. Comparisons are among deployed systems rather than isolated base models.}
\label{tab:supp-systems}
\end{table*}

%% file: tables/table_full_design.tex
\begin{table*}[t]
\centering
\begin{minipage}[t]{0.48\textwidth}
\centering
\small
\setlength{\tabcolsep}{3.2pt}
\begin{tabular}{@{}clcccc@{}}
\toprule
Release & Description & Prior & $C_{\mathrm{FA}}$ & $C_{\mathrm{FR}}$ & Threshold \\
\midrule
A & Equal-cost baseline & .50 & 1 & 1 & $1/2$ \\
B & Low-prior control & .10 & 1 & 1 & $1/2$ \\
C & High false-accept policy & .50 & 10 & 1 & $.091$ \\
D & Higher-cost dose & .50 & 20 & 1 & $.048$ \\
\bottomrule
\end{tabular}
\end{minipage}\hfill
\begin{minipage}[t]{0.48\textwidth}
\centering
\small
\setlength{\tabcolsep}{3.5pt}
\begin{tabular}{@{}lrr@{}}
\toprule
Stage & Calls/reviewer & Across four \\
\midrule
Risk-only scores & 720 & 2,880 \\
Controlled interventions & 2,188 & 8,752 \\
Operational decisions & 1,000 & 4,000 \\
Repeat measurements & 40 & 160 \\
\midrule
Total & 3,948 & 15,792 \\
\bottomrule
\end{tabular}
\end{minipage}
\caption{Prompt conditions and accepted-call counts. Condition B changes the stated prior while the action policy remains fixed. Conditions C and D change the false-accept cost and print the corresponding threshold.}
\label{tab:supp-design}
\end{table*}

%% file: tables/table_interfaces.tex
\begin{table*}[t]
\centering
\small
\renewcommand{\arraystretch}{1.08}
\begin{tabular}{@{}p{0.11\textwidth}p{0.26\textwidth}p{0.18\textwidth}p{0.20\textwidth}r@{}}
\toprule
Interface & Information shown after patch context & Scientific output & Main use & Calls/reviewer \\
\midrule
Risk-only & no card, prior policy, cost, threshold, or action request & failure probability & calibration and external fusion & 720 \\
Joint probability/action & evidence card, stated prior, cost, and threshold as defined by arm & probability and action & policy and evidence interventions & 1,440 \\
Action-only & evidence card and decision context & action & action response to evidence & 480 \\
Direct selector & both candidates and candidate-specific evidence & selected candidate & best-of-two comparison & 200 \\
Arithmetic & synthetic prior, likelihood, and cost & posterior, threshold, action & controlled arithmetic check & 60 \\
Action-then-probability & same evidence and policy; output order reversed & action then probability & elicitation-order intervention & 48 \\
\bottomrule
\end{tabular}
\caption{Response formats. Counts refer to the complete matrix per reviewer. The risk-only call is a distinct condition, not a reinterpretation of the equal-cost joint prompt.}
\label{tab:supp-interfaces}
\end{table*}

%% file: tables/table_arms.tex
\begin{table*}[t]
\centering
\small
\setlength{\tabcolsep}{5pt}
\begin{tabular}{@{}p{0.34\textwidth}p{0.47\textwidth}r@{}}
\toprule
Arm & Purpose & Calls \\
\midrule
R\_CODE\_CAL\_A & Risk-only score on CAL\_A & 160 \\
R\_CODE\_CAL\_B & Risk-only score on CAL\_B & 160 \\
R\_CODE\_EVAL & Risk-only score on EVAL & 400 \\
R\_CODE\_REPEAT & Repeated risk-only calls & 40 \\
C\_PROB\_SYM & Probability response to symmetric reliability cards & 480 \\
C\_ACTION\_SYM & Action response to symmetric reliability cards & 480 \\
C\_NEUTRAL\_LABEL\_R080 & Alternative lexical labels & 160 \\
C\_PROSE\_FORMAT\_R080 & Prose rather than tabular card format & 160 \\
C\_ASYMMETRIC\_PROB & Cards with separate sensitivity and specificity & 160 \\
C\_PROB\_THEN\_ACTION\_R080 & Equal-cost probability followed by action & 160 \\
C\_ACTION\_THEN\_PROB\_R080 & Action requested before probability & 48 \\
C\_COST\_B\_C\_R080 & Low-prior and cost-10 conditions & 320 \\
C\_COST\_D\_R080 & Cost-20 condition & 80 \\
C\_ARITHMETIC & Synthetic Bayesian arithmetic cases & 60 \\
C\_PBA\_REPEAT\_R080 & Repeated joint probability/action calls & 80 \\
O\_ACTUAL\_SETTING\_A & Operational monitor card under equal costs & 400 \\
O\_ACTUAL\_SETTING\_C & Operational monitor card under cost 10 & 400 \\
O\_DIRECT\_SELECTOR & Direct two-candidate comparison & 200 \\
\bottomrule
\end{tabular}
\caption{Per-reviewer call matrix. Counts sum to 3,948. CAL and EVAL are repository-disjoint splits. Controlled panels are fixed subsets defined without reviewer outcomes.}
\label{tab:supp-arms}
\end{table*}

%% file: tables/table_prompt_audit.tex
\begin{table}[b]
\centering
\small
\setlength{\tabcolsep}{3pt}
\begin{tabular}{@{}lrrr@{}}
\toprule
Comparison & Pairs & Same system prompt & Only policy lines differ \\
\midrule
A--C & 640 & 640 & 640 \\
A--D & 320 & 320 & 320 \\
C--D & 320 & 320 & 320 \\
\bottomrule
\end{tabular}
\caption{Controlled prompt audit. Counts are reviewer--candidate--signal pairs.}
\label{tab:supp-prompts}
\end{table}

%% file: tables/table_prior_control.tex
\begin{table*}[t]
\centering
\small
\setlength{\tabcolsep}{5.2pt}
\begin{tabular}{@{}lcc@{}}
\toprule
Reviewer & Prior .10 minus .50 & Cost-10 policy minus equal cost \\
\midrule
DeepSeek & $-0.204$ [$-0.237$,$-0.168$] & $-0.068$ [$-0.111$,$-0.024$] \\
Grok & $-0.239$ [$-0.279$,$-0.195$] & $+0.001$ [$-0.028$,$+0.028$] \\
Mistral & $-0.094$ [$-0.112$,$-0.076$] & $-0.079$ [$-0.102$,$-0.057$] \\
GPT/Codex & $-0.179$ [$-0.207$,$-0.149$] & $-0.096$ [$-0.125$,$-0.069$] \\
\bottomrule
\end{tabular}
\caption{Matched controlled A/B/C panel. Entries are mean changes in reported failure probability with repository-bootstrap 95\% intervals. The prior control moves all four systems in the expected direction. The signed policy effect is heterogeneous; the operational result is policy dependence, not a common conservative shift.}
\label{tab:supp-prior}
\end{table*}

%% file: tables/table_operational.tex
\begin{table*}[t]
\centering
\small
\begin{tabular}{@{}lrrrrr@{}}
\toprule
Reviewer & Brier A & Brier C & Loss A & Loss C & Rule agreement C \\
\midrule
DeepSeek & 0.255 & 0.273 & 0.965 & 2.550 & 73.8\% \\
Grok & 0.238 & 0.268 & 0.805 & 1.365 & 98.2\% \\
Mistral & 0.262 & 0.308 & 0.840 & 2.480 & 88.3\% \\
GPT/Codex & 0.232 & 0.268 & 0.705 & 3.740 & 96.8\% \\
\bottomrule
\end{tabular}
\caption{Operational behavior on EVAL under equal costs (A) and cost 10 (C). Rule agreement is the fraction of cost-10 actions that match deterministic thresholding of the reported probability.}
\label{tab:supp-operational}
\end{table*}

%% file: tables/table_repeatability.tex
\begin{table}[t]
\centering
\small
\setlength{\tabcolsep}{3pt}
\begin{tabular}{@{}lrrl@{}}
\toprule
Reviewer & $|C-A|$ & Repeat movement & Excess [95\% CI] \\
\midrule
DeepSeek & 0.174 & 0.154 & 0.020 [-0.027, 0.066] \\
Grok & 0.090 & 0.052 & 0.038 [0.013, 0.066] \\
Mistral & 0.114 & 0.046 & 0.068 [0.047, 0.095] \\
GPT/Codex & 0.154 & 0.105 & 0.049 [0.029, 0.069] \\
\bottomrule
\end{tabular}
\caption{Movement caused by the policy block compared with movement between repeated calls under the same prompt.}
\label{tab:supp-repeat}
\end{table}

%% file: tables/table_thresholds.tex
\begin{table}[t]
\centering
\small
\begin{tabular}{@{}lrr@{}}
\toprule
Reviewer & Cost 10 boundary & Cost 20 boundary \\
\midrule
DeepSeek & 1/80 & 4/40 \\
Grok & 0/80 & 0/40 \\
Mistral & 37/80 & 18/40 \\
GPT/Codex & 0/80 & 0/40 \\
\bottomrule
\end{tabular}
\caption{Exact reports of the printed threshold under CLEAR evidence.}
\label{tab:supp-threshold}
\end{table}

%% file: tables/table_stage_inference.tex
\begin{table*}[t]
\centering
\small
\setlength{\tabcolsep}{5.0pt}
\begin{tabular}{@{}lccc@{}}
\toprule
Reviewer & Cost-10 $p$ $-$ equal-cost $p$ & Model action $-$ cost-10 $p$ & Model action $-$ equal-cost $p$ \\
\midrule
DeepSeek & 1.270 [0.724,1.816] & 0.150 [-0.576,0.883] & 1.420 [0.925,1.945] \\
Grok & 0.235 [0.037,0.464] & 0.130 [-0.020,0.322] & 0.365 [0.117,0.654] \\
Mistral & 1.430 [0.934,1.943] & 0.040 [-0.470,0.546] & 1.470 [0.932,2.021] \\
GPT/Codex & 2.450 [1.840,3.072] & 0.210 [0.017,0.429] & 2.660 [2.000,3.350] \\
\bottomrule
\end{tabular}
\caption{Sources of cost-10 loss. Entries are loss differences per issue with repository-bootstrap 95\% intervals. Deterministic policies use the exact decimal threshold printed in the prompt. Cost-10 probability minus equal-cost probability changes only the probability source under the same decision rule.}
\label{tab:supp-stage}
\end{table*}

%% file: tables/table_precision_audit.tex
\begin{table*}[t]
\centering
\small
\setlength{\tabcolsep}{5pt}
\begin{tabular}{@{}lrr@{}}
\toprule
Reviewer & Reports at displayed boundary & Decisions changed by binary-float comparison \\
\midrule
DeepSeek & 4 & 4 \\
Grok & 0 & 0 \\
Mistral & 55 & 66 \\
GPT/Codex & 0 & 0 \\
\bottomrule
\end{tabular}
\caption{Threshold-precision audit. Paper results compare the canonical decimal report with the exact decimal string shown in the prompt. A binary floating-point approximation can classify displayed-boundary reports incorrectly.}
\label{tab:supp-precision}
\end{table*}

%% file: tables/table_modular_controller.tex
\begin{table*}[t]
\centering
\small
\setlength{\tabcolsep}{4.0pt}
\begin{tabular}{@{}llrrrr@{}}
\toprule
& & \multicolumn{2}{c}{Equal cost} & \multicolumn{2}{c}{Cost 10} \\
\cmidrule(lr){3-4}\cmidrule(l){5-6}
Reviewer & Controller score & Loss & Accept (\%) & Loss & Accept (\%) \\
\midrule
DeepSeek & Risk only & 0.985 & 37.8 & 1.000 & 0.0 \\
DeepSeek & Monitor only & 0.830 & 67.5 & 1.000 & 0.0 \\
DeepSeek & Binary-card fusion & 0.840 & 52.0 & 1.000 & 0.0 \\
DeepSeek & Continuous fusion & 0.800 & 64.0 & 1.000 & 0.0 \\
Grok & Risk only & 0.860 & 57.0 & 1.000 & 0.0 \\
Grok & Monitor only & 0.830 & 67.5 & 1.000 & 0.0 \\
Grok & Binary-card fusion & 0.850 & 57.0 & 1.000 & 0.0 \\
Grok & Continuous fusion & 0.820 & 62.0 & 1.000 & 0.0 \\
Mistral & Risk only & 0.910 & 58.0 & 1.000 & 0.0 \\
Mistral & Monitor only & 0.830 & 67.5 & 1.000 & 0.0 \\
Mistral & Binary-card fusion & 0.815 & 52.8 & 1.000 & 0.0 \\
Mistral & Continuous fusion & 0.850 & 67.5 & 1.000 & 0.0 \\
GPT/Codex & Risk only & 0.695 & 59.2 & 1.000 & 0.0 \\
GPT/Codex & Monitor only & 0.830 & 67.5 & 1.000 & 0.0 \\
GPT/Codex & Binary-card fusion & 0.705 & 59.2 & 1.000 & 0.0 \\
GPT/Codex & Continuous fusion & 0.690 & 57.5 & 1.000 & 0.0 \\
\bottomrule
\end{tabular}
\caption{Risk-only pipeline at equal costs and cost 10. Scores come from risk-only reviewer calls or the independent monitor measurement, are fitted on pooled CAL, and are frozen before EVAL. The binary-card model uses the same WARNING/CLEAR representation shown in the joint prompts. Every variant matches reject-all at cost 10.}
\label{tab:supp-separate-controller}
\end{table*}

%% file: tables/table_nested_refit.tex
\begin{table}[t]
\centering
\small
\setlength{\tabcolsep}{4.0pt}
\begin{tabular}{@{}lrrr@{}}
\toprule
Reviewer & Mean gain & 95\% interval & Positive draws (\%) \\
\midrule
DeepSeek & 0.0156 & [0.0067,0.0253] & 100.0 \\
Grok & 0.0088 & [0.0021,0.0169] & 99.4 \\
Mistral & 0.0146 & [0.0065,0.0242] & 100.0 \\
GPT/Codex & 0.0012 & [-0.0033,0.0057] & 79.4 \\
Equal cohort & 0.0100 & [0.0039,0.0175] & 99.7 \\
\bottomrule
\end{tabular}
\caption{Nested refit sensitivity for the Brier-score gain from combining reviewer and monitor scores. Each of 1,000 draws resamples CAL repositories, refits both models, and independently resamples EVAL repositories. Positive values favor the combined model.}
\label{tab:supp-nested}
\end{table}

%% file: tables/table_monitor.tex
\begin{table*}[t]
\centering
\small
\setlength{\tabcolsep}{5pt}
\begin{tabular}{@{}lrrrrrr@{}}
\toprule
Reviewer & Risk-only Brier & Monitor-only Brier & Additive Brier & Interaction Brier & Risk pair (\%) & Additive pair (\%) \\
\midrule
DeepSeek & 0.2496 & \textbf{0.2325} & 0.2334 & 0.2332 & 54.8 & 63.0 \\
Grok & 0.2420 & \textbf{0.2325} & 0.2328 & 0.2326 & 63.5 & 63.2 \\
Mistral & 0.2486 & \textbf{0.2325} & 0.2336 & 0.2338 & 54.5 & 59.5 \\
GPT/Codex & 0.2216 & 0.2325 & \textbf{0.2199} & 0.2198 & 73.5 & 70.8 \\
\bottomrule
\end{tabular}
\caption{Reviewer and monitor scores on EVAL. Lower Brier is better; bold marks the lowest of reviewer-only, monitor-only, and combined point estimates in each row. Pair accuracy gives half credit to exact ties. All fitting and model selection use CAL only.}
\label{tab:supp-monitor}
\end{table*}

%% file: tables/table_reliability_cells.tex
\begin{table}[t]
\centering
\small
\setlength{\tabcolsep}{4.2pt}
\begin{tabular}{@{}lrrr@{}}
\toprule
Reviewer & $r=.55$ & $r=.80$ & $r=.95$ \\
\midrule
DeepSeek & 0.0127 & 0.0552 & 0.2119 \\
Grok & 0.0059 & 0.0542 & 0.0933 \\
Mistral & 0.0032 & 0.0022 & 0.0033 \\
GPT/Codex & $-0.0032$ & 0.0455 & 0.0088 \\
\bottomrule
\end{tabular}
\caption{Reduction in expected Brier score from applying the card likelihood ratio in code. Each cell compares the coded update with the reviewer's in-prompt update after receiving the same card and stated reliability. Positive values favor the coded update; 11 of 12 cells are positive.}
\label{tab:supp-reliability-cells}
\end{table}

%% file: tables/table_pairwise.tex
\begin{table}[t]
\centering
\small
\setlength{\tabcolsep}{2.5pt}
\begin{tabular}{@{}lrrr@{}}
\toprule
Reviewer & Base $\rightarrow$ cascade & Gain pp [95\% CI] & Esc. (\%) \\
\midrule
DeepSeek & 54.8 $\rightarrow$ 57.5 & 2.75 [0.53, 5.00] & 10.5 \\
Grok & 63.5 $\rightarrow$ 66.0 & 2.50 [0.27, 4.70] & 10.0 \\
Mistral & 54.5 $\rightarrow$ 59.0 & 4.50 [0.44, 8.41] & 31.0 \\
GPT/Codex & 73.5 $\rightarrow$ 76.0 & 2.50 [0.50, 4.57] & 9.0 \\
\bottomrule
\end{tabular}
\caption{Best-of-two collision cascade. This is a candidate-selection result, not a one-patch approval result.}
\label{tab:supp-pairwise}
\end{table}